# Generation and Optimization of Test cases for Object-Oriented Software Using State Chart Diagram

Ranjita Kumari Swain[1], Prafulla Kumar Behera[2] and Durga Prasad Mohapatra[3]

[1]Rourkela Institute of Mgt. Studies, Rourkela
ranjita762001@yahoo.com
[2]Dept. of Comp. Sc., Utkal University, Bhubaneswar
P_behera@hotmail.com
[3]Dept. of Comp. Sc. and Engg., National Institute of Technology, Rourkela
durga@nitrkl.ac.in

## ABSTRACT

*The process of testing any software system is an enormous task which is time consuming and costly. The time and required effort to do sufficient testing grow, as the size and complexity of the software grows, which may cause overrun of the project budget, delay in the development of software system or some test cases may not be covered. During SDLC (software development life cycle), generally the software testing phase takes around 40-70% of the time and cost. State-based testing is frequently used in software testing. Test data generation is one of the key issues in software testing. A properly generated test suite may not only locate the errors in a software system, but also help in reducing the high cost associated with software testing. It is often desired that test data in the form of test sequences within a test suite can be automatically generated to achieve required test coverage.*

*This paper proposes an optimization approach to test data generation for the state-based software testing. In this paper, first state transition graph is derived from state chart diagram. Then, all the required information are extracted from the state chart diagram. Then, test cases are generated. Lastly, a set of test cases are minimized by calculating the node coverage for each test case. It is also determined that which test cases are covered by other test cases. The advantage of our test generation technique is that it optimizes test coverage by minimizing time and cost. The proposed test data generation scheme generates test cases which satisfy transition path coverage criteria, path coverage criteria and action coverage criteria. A case study on Automatic Ticket Machine ( ATM ) has been presented to illustrate our approach.*

## 1. INTRODUCTION

System testing is the most complex and intricate among all types of program testing. The complexity of system testing can possibly be attributed to the fact that it involves testing a fully integrated system that may be large and complex. Therefore, with continually increasing system sizes, the issue of automatic design of system test cases is assuming prime importance [26]. A properly generated test suite may not only locate the errors in a software system, but also help in reducing the high cost associated with software testing [16]. Many present day software solutions are state based. In such systems, the system behavior is determined by its state. In other words, a system can respond differently to the same event in different states. Therefore, unless a system is made to assume all its possible states and tested, it would not be possible to uncover state-based bugs. Adequate system testing of such software requires satisfactory coverage of system states and transitions. Generation of test specifications to meet these coverage criteria can be accomplished by using the state model of a system. However, it is a non-trivial task to manually construct the state model of a system. The state model of an actual system is usually extremely complex and comprises of a large number of states and transitions. Possibly for this reason, state models of complete systems are rarely constructed by system developers [26].

Testing activities consist of designing test cases that are sequences of inputs, executing the program with test cases and examining the results produced by this execution. Testing can be carried out earlier in the development process so that the developer will be able to find the inconsistencies and ambiguities in the specification and hence will be able to improve the specification before the program is written [11]. Unified modeling language (UML) has emerged as the de facto standard for modeling software systems and has received significant attention from researchers as well as practitioners. UML models are popular not only for designing and documenting systems, the importance of UML models in designing test cases has also been well recognized. Even though UML models are intended to help reduce the complexity of a problem, with the increase in product sizes and complexities, the UML models themselves become large and complex involving thousands of interactions across hundreds of objects [15]. The important part of quality control in the software life-cycle is testing. As the complexity and size of software increase, the time and effort required to do sufficient testing grow. Manual testing is time-consuming and error prone. So, there is a pressing to automate the testing process. The testing process can be divided into three parts: test case generation, test execution, and test evaluation. The latter two parts are relatively easy to automate provided that the criteria for passing the tests are available. However, to determine which tests are required to achieve a certain level of confidence is not trivial [17].

Model-based testing [5] has grown in importance. Models are specified to represent the relevant, desirable features of the system under consideration (SUC). These models are used as a basis for (automatically) generating test cases to be applied to the SUC. Typical models that are used for representing system behavior are unified modeling language, finite state machines, state charts etc. [3].

With this motivation, we aim our work at deriving the test sequence from state transition diagram and maximizing state or node coverage. Also our method minimizes the size of test, time and cost, while preserving test coverage. The rest of the paper is structured as follows: A brief discussion on UML diagrams, which are relevant to our paper is described in the Section 2. Then, we discuss some testing coverage criteria in Section 3. Section 4 represents some concepts, notations and definitions of state chart diagram. In Section 5, we explain the overview of our proposed method for construction of state transition graph, generation of test sequence using state charts and how node coverage is calculated for each test case. Section 6 provides the working of our methodology with the ATM (Automatic Ticket Machine) case study. Section 7 discusses some related work. Finally, Section 7 presents the conclusion and future work of this paper.

## 2. UML DIAGRAMS

In this section, we discuss some basic concepts which will be used subsequently in our paper. UML, Unified Modeling Language is a visual language that has been developed to support the design of complex object-oriented systems. Since its introduction in the late 90s, it has undergone several revisions. The latest release being UML version 2.0, which adds several new capabilities to UML 1.x. In this section, we restrict our review to only those developments that are directly relevant to our work. A typical software procedure incorporates all the three aspects: It uses data structure (class model), it sequences operations in time (state model), and it passes data and control among objects (interaction model). The three kinds of models separate a system into different views. The different models are not completely independent but a system is more than a collection of these independent parts. UML specification defines two major kinds of UML diagram: *structural diagrams* and *behavioral diagrams*.

The elements in a structural diagram represent the meaningful concepts of a system, and may include abstract, real world and implementation concepts. Behavioral diagrams show the dynamic behavior of the objects in a system, which can be described as a series of changes to the system over time.

## 2.1. UML state chart diagram

In this section, we explain the few fundamentals on state chart diagram. The name of the diagram itself clarifies the purpose of the diagram and other details. It describes different states of a component in a system. The states are specific to a component or object of a system. A state chart diagram describes a state machine. Now to clarify it state machine can be defined as a machine which defines different states of an object and these states are controlled by external or internal events. State chart diagram is one of the five UML diagrams used to model dynamic nature of a system. They define different states of an object during its lifetime. And these states are changed by events. So state chart diagrams are useful to model reactive systems. Reactive systems can be defined as a system that responds to external or internal events. State chart diagram describes the flow of control from one state to another state. States are defined as a condition in which an object exists and it changes when some event is triggered. So the most important purpose of State chart diagram is to model life time of an object from creation to termination. State chart diagrams are also used for forward and reverse engineering of a system. But the main purpose is to model reactive system.

State diagrams are used to give an abstract description of the behavior of a system. This behavior is analyzed and represented in series of events that could occur in one or more possible states. Hereby "each diagram usually represents objects of a single class and tracks the different states of its objects through the system". State diagrams can be used to graphically represent finite state machines. Followings are the main purposes of using state chart diagrams:

- To model dynamic aspect of a system.
- To model life time of a reactive system.
- To describe different states of an object during its lifetime.
- To define a state machine to model states of an object.

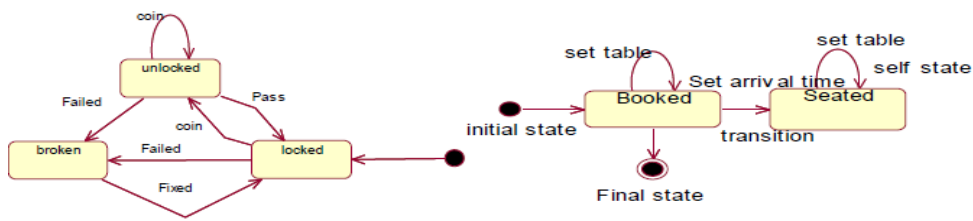

Figure. 1. Basic concepts of a simple state chart for a coin machine and reservation class

The basic components of a state chart diagram are - events, states and transitions. But it has some distinguishing characteristics for modeling dynamic nature. State chart diagram defines the states of a component and these state changes are dynamic in nature. So, its specific purpose is to represent the state changes triggered by events. Events are internal or external factors influencing the system. If we look into the practical implementation of state chart diagram then it is mainly used to analyze the object states influenced by events. This analysis is helpful to understand the system behavior during its execution. The following are the basic notational elements that can be used to make up a diagram:

**2.1.1. State:** The state of an object is represented by rectangle with rounded corners. Top of the rectangle contains a name of the state. It can contain a horizontal line in the middle, below which the activities that are done in that state are indicated. A state in a state chart diagram can either be simple or composite type. A simple state does not have any sub-states. A composite state, on the other hand, consists of one or more regions. A region is a container for sub-states.

**2.1.2. Initial state:** A transition leading from an initial event shows the state that an object goes into when it is created or initialized. This is shown as a small black disk or filled circle. A state-chart can have only one initial state.

**2.1.3. Final state:** Like initial state the state diagram shows final state. It represents the state reached when an object is destroyed, switched off or stops responding to events. This is shown as a hollow circle containing a smaller filled circle or small black disk within a large circle. A state-chart may have more than one final state. A transition to a history state causes the sub state that was most recently active in the composite state to become active again [22].

**2.1.4. Transition:** Arrows, denote transitions. The name of the event (if any) causing this transition is written as the labels with the transition names or event names. A guard expression may be added before a "/" and enclosed in square-brackets (event Name [guard Expression]), denoting that this expression must be true for the transition to take place. The general format of the label of a transition is as follows:

Event [guard-constraint] '/' activity-expression/action

**2.1.5. An event:** An event is an occurrence at a point of time. Events often correspond to verbs in the past tense e.g. (power turned on, alarm set) [4]. Events include error conditions as well as normal occurrences, e.g. motor jammed, transaction aborted etc.

## 3. TEST ADEQUECY CRITERIA

Testing coverage / adequacy criterion specifies the requirement of a particular testing, and can be used as an objective measurement of the test case. In traditional software code testing, the definition of testing adequacy is defined as a measurement function. The case of UML state chart diagrams is different because it is in the form of model instead of code. Especially the coverage of state chart diagram is little bit complex. In this paper, we propose different types of coverage metrics as follows:

- *State Coverage* requires that all the state nodes in a state chart diagram be covered. The value of state coverage is the ratio between the covered states and all the states in the state chart diagram.

- *Action Coverage*: In order to generate test cases with respect to action coverage we construct as many marked specifications as there are actions within the specification.

- *Transition Coverage* requires that all the completion transitions in a state chart diagram be covered. The value of transition coverage is the ratio between the checked transitions and all the transitions in the state chart diagram.

- *Path Coverage* requires that all the paths in a state chart diagram be covered. The value of path coverage is the ratio between the traversed paths and all the paths in the diagram.

- *Condition coverage*. A decision consists of conditions separated by logical operators (e.g. and, or). A single condition is covered, if it evaluates to both true and false at some point during test execution. Decision coverage has also been called branch coverage or predicate coverage. This means that 100% condition / decision coverage is achieved if all conditions evaluate to true and to false and if every decision also evaluates to true and to false during the test execution.

## 4. SOME BASIC DEFINITIONS

There are many forms of state diagrams, which differ slightly and have different semantics. For a deterministic finite state machine (DFA), nondeterministic finite state machine (NFA), generalized nondeterministic finite state machine (GNFA), Mealy machine or Moore machine, the input is denoted on each edge. For a Mealy machine, input and output are signified on each edge, separated with a slash "/": "1/0" denotes the state change upon encountering the symbol "1" causing the symbol "0" to be output. For a Moore machine the state's output is usually written inside the state's circle, also separated from the state's designator with a slash "/". There are also variants that combine these two notations.

It is assumed that a state chart STc is correct in the sense that for each state $s \in S_{simple}$ there exists a sequence of transitions $t_1, t_2..., t_k$ so that $source(t_1) \in Si$ and $target(t_k) = s$ and for each state $s \in S_{simple}$ there exists a sequence of transitions $t_1, t_2..., t_k$ so that $source(t_1) = s$ and $target(t_k) \in S_f$.

The following terms will be used to describe our technique.

*Definition 1.* A **state chart** can be a quadruple $S_c$ = (E, $S_t$, H, T), where E is a finite set of events and $St$ = (S, $S_i$, $S_f$) is a triple of set of states with S as a finite set of states, $S_i \subseteq S$ denoting the entries (initial states) and $S_f \subseteq S$ the exits (final states),

H $\subseteq$ S $\times$ S is a hierarchy relation, a binary relation on the set S forming a tree. For an element (s, s′) $\in$ H holds, that a state s is an immediate sub state of state S′.

T $\subseteq$ S $\times$ E $\times$ S is a finite set T of transitions. The set of states S is composed of disjoint sets of simple states $S_{simple}$ and composite states $S_{comp}$.

*Definition 2.* A **transition pair** TP = (t, t′) with t, t′ belongs to $T_{legal}$ is a sequence of a legal incoming transition to a legal outgoing transition of a (simple) state so that $\exists s \in S_{simple}$:

t $\in$ in(s) U t′ $\in$ out(s).

*Definition 3.* A **false transition pair** FP = (t, t′) with t $\in T_{legal}$ and t′ $\in T_{faulty}$ is a sequence of a legal incoming transition to a faulty outgoing transition of a (simple) state so that $\exists s \in S_{simple}$ : t $\in$ in(s) U t′ $\in$ out(s).

*Definition 4.* A sequence of n legal transitions $(t_1, t_2..., t_n)$ with $t_i \in T_{legal}$ where $(t_i, t_{i+1})$ denotes a valid transition pair for all i $\in$ 1, ..., n - 1 is called a **transition sequence** ($T_{seq}$) of length n. A transition sequence $(t_1, t_2..., t_n)$ is complete if it starts at the initial state of the state chart that is entered firstly and ends at a final state. In this case it is called a complete transition sequence ($T_{sqcom}$).

*Definition 5.* A **fault transition sequence** $T_{sqfault} = (t_1, t_2..., t_n)$ of length n consists of n - 1 subsequent transitions, forming a (legal) transition sequence of the length n - 1 plus a concluding, faulty transition $t_n \in T_{faulty}$. A faulty transition sequence is called complete if it starts at the initial state of the state chart, abbreviated as CFTS. The sequence ($t_1, t_2..., t_{n-1}$) is called a start sequence.

*Definition 6.* A **test case** is the triplet [I, S, O], where I is the initial state of the system at which the test data is input, S is the test data input to the system and O is the expected output of the system [18], [19]. The output produced by the execution of the software with a particular test case provides a specification of the actual software behavior. A test case is also characterized by an ordered pair of an input and an expected output of the SUC.

*Definition 7:* A **test suite** is a set of test cases. A single test case in most cases may satisfy more than one test obligation. For instance, a test case used to cover a certain state of interest may also cover other states during its execution. This then provides for a way to reduce the size of the final test suite by choosing a subset of test cases that preserves the coverage obtained by the full test suite [9]. Based on Definitions 4 and 5 the following two coverage criteria are introduced:

*Definition 8:* **k-transition coverage** (k-TC) generates complete transition sequences that sequentially conduct all legal transition sequences of length $k \in N$.

*Definition 9:* Faulty transition pair coverage (FTC), generates complete faulty transition sequence for each faulty transition and faulty transition pair[13].

Definition 8 guarantees that all possible (legal) transition sequences of length k will be tested. A test suite consisting of all transition sequences of a fixed length k does not necessarily cover a set of all sequences of length $i \in 1, ...k-1$ as there may exist sequences of length i that cannot be expanded to length k. Definition 9 guarantees that all potential malfunctions will be tested. Typically, state based test generation methods focus on some form of coverage, for instance on covering transitions [19], [20] or on transition coverage and state identification [6], [14]. Our approach creates all transition sequences of length k including all shorter sequences of length 1, ..., k -1 that cannot be found in longer sequences.

## 5. GeMiTefSc–OUR PROPOSED APPROACH TO GENERATE AND MINIMIZE TEST SEQUENCE FROM STATE CHART DIAGRAM

In this section, we discuss a overview of our proposed approach to generate test sequence from UML state chart diagram and then, we optimize the test node coverage while minimizing time and cost. We have named our approach, *Generation and Minimization of test cases from State Charts* (GeMiTefSc).

Our approach consists of 5 steps. The schematic representation of our approach is shown in fig.2.

- Step 1: Building model of the system.

- Step 2: Constructing the transition graph from state chart diagram.

- Step 3: Extracting the required information from the transition.

- Step 4: Generating test cases.

- Step 5: Minimize a set of test cases by calculating node coverage for each test sequence.

In the next section, we discuss each step in detail, by using different algorithms for each step in different sections. Before that we present some related definitions.

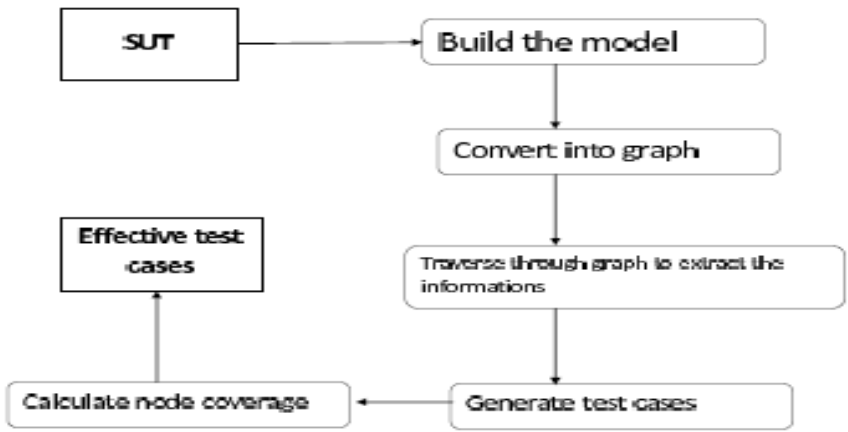

Fig. 2. Schematic representation of our approach

## 5.1. Building model of the system.

State diagram is graph where nodes represent states end the directed arcs that interconnect states represent transitions. It also models dynamic behavior, and captures the different states that an object can be in, and its response to various events that may arise in each of its states. State chart diagram is one of the 13 UML diagrams used to model dynamic nature of a system. They define different states of an object during its lifetime. The notation and semantics of UML state diagrams are substantially based on State charts modified to include object-oriented features [12]. The states and the transition of a system are important to set up a state diagrams from system. Fig. 3 shows a UML state diagram for a ticket machine system.

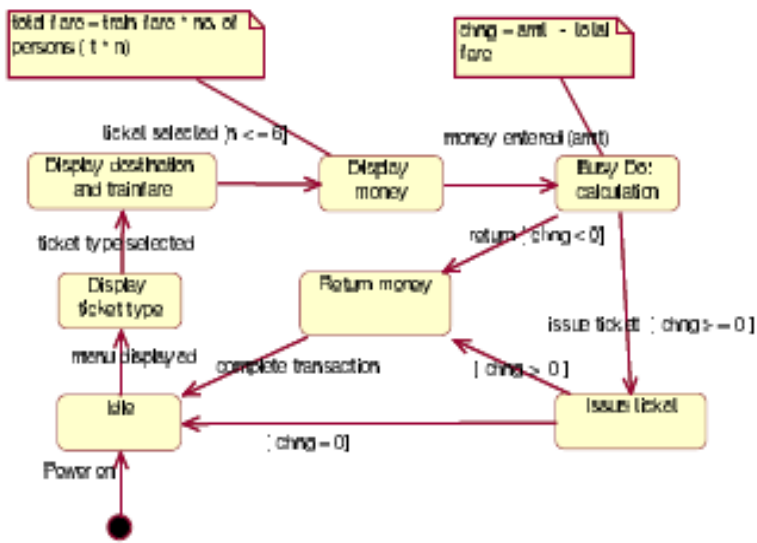

Figure. 3. State chart diagram for Automated Ticket Machine

## 5.2. Constructing the transition graph from state chart diagram

Here, we convert the state diagram into state transition graph. A state transition graph

TG = ($V_t$, $E_d$).

***Definition 10:*** A **transition graph** TG = ($V_t$, $E_d$) represents a directed graph consisting of a set of vertices ($V_t$), a set of directed edges ($E_d$). In TG, nodes represent states and edges represent transitions between states. Without any loss of generality, we assume that there is a unique node that corresponds to the initial state and that one or more nodes represent the final states. The initial state is represented as the root of the tree. States at each level of nesting are considered as a sub graph. We represent this step through an algorithm.

**Algorithm-1**

*Input: Sc* = **(E, $S_t$, H, T)**
*Output*: **A transition graph TG = ($V_t$, $E_d$)**

1.     V := {$t_i$, $t_j$}, A := NULL
2.     **Foreach** t ∈ $T_{legal}$ do
3.         V := V U {t}
4.         **Foreach** s ∈ $S_{simple}$ do
5.             **Foreach** t ∈ in(s) do
6.                 **Foreach** t′ ∈ out(s) $T_{faulty}$ do
7.                       A := A U {t,t′ }
8.                 **Endfor**
9.             **Endfor**
10.         **Endfor**
11.         **Foreach** t ∈ out(initial(root())) $T_{faulty}$ do
12.             A := A U {$t_i$,t}
13.         **Endfor**
14.         **Foreach** s ∈ $S_f$ do
15.             **Foreach** t ∈ in(s) do
16.                 A := A U {t, $t_j$}
17.             **Endfor**
18.         **Endfor**

Finally, the transition graph is used as a source for the graphical traveling salesman problem and it is augmented by an additional edge ($t_i$, $t_j$). This edge and the preconditions for a state chart claimed in the last section ensure that the resulting graph is strongly connected.

### 5.3. Extracting all required information from the graph

Here, in this subsection, we present how to extract all the information which are required to generate the test sequences.

*Input: Transition graph TG = ($V_t$, $E_d$)*

*Output*: A set of stages or node ST, set of input data ID, set of output data OD and set of transition TR.

ST = {St1, St2, St3 ....Stn}, each Sti is a stage or node

ID = {ID1, ID2, ID3 ....IDn}, each IDi is an input data

OD = {OD1, OD2, OD3 ....ODn}, each ODi is an output value TR = {TR1, TR2, TR3 ....TRn}, each TRi is a transition between source and destination stage, where each TRi = {Stp, Stq}, Stp is a source stage and Stq is a destination stage . Algorithm - 1 describes how a transition graph is constructed for a state chart $Sc = (E, St, H, T)$ : Set V = { t/ t ∈ $T_{legal}$} consists of vertices representing the legal transitions of state chart Sc. For each (legal) transition pair (t, t′ ) of the state chart, a directed edge is created. Vertex $t_i$ has to be connected with all transitions that may be triggered from the state belonging to the initial configuration. Transitions leading into a final state have to be connected with vertex $t_j$ .

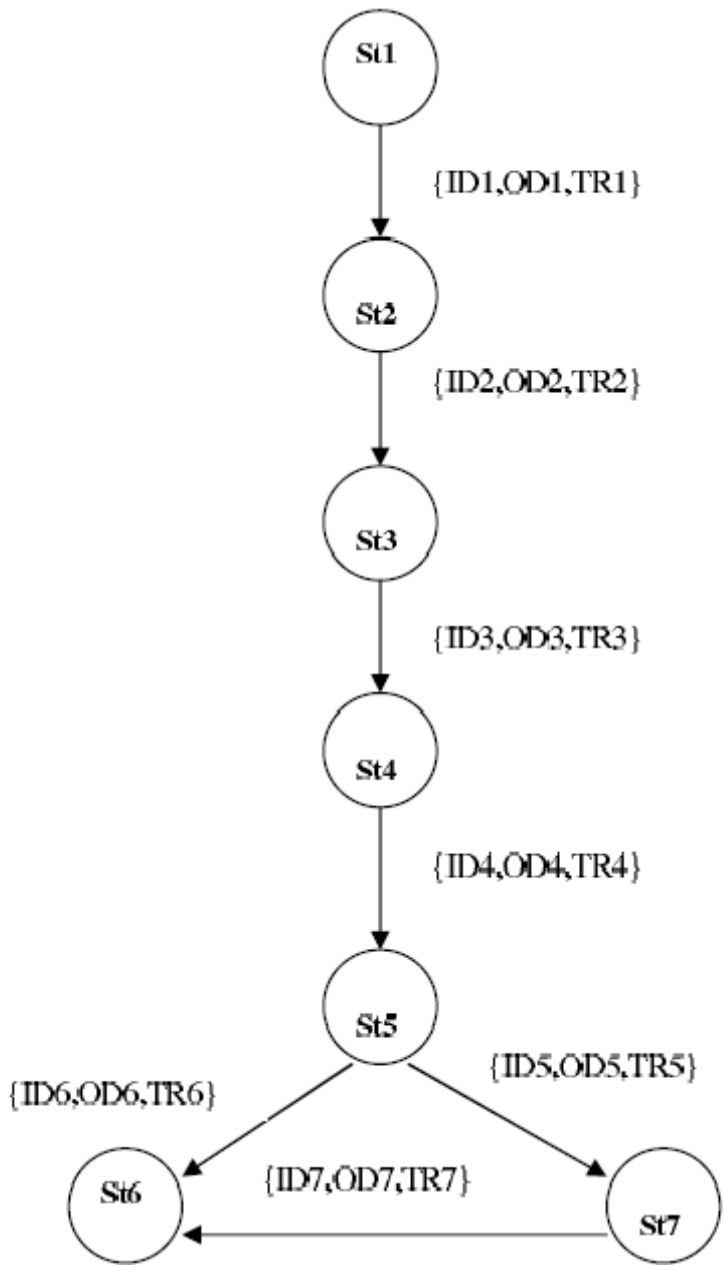

Figure 4. Corresponding transition graph of the state chart diagram of ATM

A simple transition coverage may be reached by visiting all vertices of the graph, based on a transition graph at least once by starting in vertex $t_i$ and ending in vertex $t_j$. The problem of computing a route for visiting all vertices of a graph by minimizing the length of the route is well known as the traveling salesman problem (TSP). If visiting vertices and traversing edges more than once is allowed, it is called the graphical traveling salesman problem (GTSP) [23]. A test case set fulfilling k-transition coverage for k > 1 is computed by transforming the transition graph stepwise and then applying the graphical traveling salesman problem. Also by computing all complete transition sequences whose length is smaller than k and that cannot be expanded to longer sequences, a minimal test case set fulfilling k-transition coverage for all k $\in$ 1, ..., n is achieved. This proceeding is described in Algorithm-2, in the next section.

### 5.4. Generating test cases

After, a transition graph is constructed based on a state chart *Sc*. If n = 1, the graphical traveling salesman problem can be applied directly. If n is greater than 1 the transition graph has to be transformed k - 1 times. The resulting graph represents all possible sequences of transitions of length k. Additionally, all sequences of length k-1 are computed that cannot be expanded to longer sequences. These sequences are characterized by the fact that the corresponding vertex representing that sequence is solely connected with vertices $t_i$ and $t_j$. The functions indeg(v) := v′,(v′, v) $\in$ A and outdeg(v) := v′(v, v′) $\in$ A are used to compute these vertices. As these sequences are already complete, they can be added to the set TS.

**Algorithm-2: Generation of a test case set for k-transition coverage for k $\in$ 1, ..., n**

**Input: A state transition graph from *Sc* = = (E,*St*,H, T)), n $\in$ N**
**Output: A test case set TS fulfilling k-transition coverage**

1. **For** k $\in$ 1, ..., n
2. **Begin**
3.     i := 0; j := 1; ts := null; s := null; visited NSt := 0
4.     **Do while** NSt [i] = NULL
5.     push ( NSt [i], s);
6.     visited NSt [ NSt [i] ] := visited NSt [ NSt [i] ] + 1;
7.       **Do while** s *not* = NULL
8.         t := pop(s); NSt[i] := pop(s);
9.         **If** enabled ( NSt[i]) not = NULL
10.           visitnexttransition( NSt [i] );
11.         **Endif**
12.       **Enddo**
13.       Create a transition graph TG = (*Vt*, *Ed*) from *Sc* by applying Algorithm-1
14.       TS := $\phi$
15.     **For** k := 2 to n do
16.       **Foreach** v $\in$ V do
17.         **If** ($t_i$, v) $\in$ *Ed* ∧ (v, $t_j$) $\in$ *Ed* ∧ indeg(v) = outdeg(v) = 1 then
18.           TS := TS U v
19.           Ed := Ed U ($t_i$, $t_j$)
20.           Apply the graphical traveling salesman problem.
21.           Add them to set TS
22.         **Endif**
23.       **Endfor**
24.     **Enddo**
25.   **Enddo**
26. **End**

**27. Endfor**

## 5.5. Minimizing the test cases

Here, in this section , we discuss how the generated test cases are reduced while maximizing test coverage. Though Wang's algorithm [17] is widely used, but it does not cover other critical attributes, like defect id, dependency and automated test case indicator. In order to generate an effective size of generated test cases. This step contains two subactivities, which are

- calculate node coverage for each test case. Let $NC(tc) = t_1, t_2,...t_n$ for $NC(tc)$ to be a set of test cases that tc is covered by $t_1, t_2,...t_n$. Hence, if a number of set tc is zero, then tc is included in the effective set of test cases.

- select effective test cases. Now, we present this step through an algorithm.

**<u>Algorithm-3</u>**

**Input: *Sc* = (E,*St*,H, T) and A set of test cases TS[n];**
**Output: Reduced test suite RTs;**
1. ANC, PNC: Node coverage, Previous coverage, tc: test case
2. RTs = ϕ;
3. NC(tc) = $t_1, t_2,...t_n$
4. ANC = 0;
5. repeat
6. select a test case F from TS[n];
7. **Foreach** s ∈ $S_{simple}$ do
8.     **Foreach** t ∈ in(s) do
9.         **Foreach** t′ ∈ out(s) ∩ $T_{faulty}$ do
10.             ANC := ANC U (t, t′)
11.             **Foreach** (t, t′) ∈ ANC do
12.                 ($t_1$, ..., $t_i$) := STARTSEQUENCE(t)
13.                 RTs := RTs U ($t_1$, ...$t_i$, t, t′ )
14.                 **If** in(inital(root())) = ϕ then
15.                     RTs := RTs U (t)
16.                     select effective test cases;
17.                     **If** a number of set tc is zero;
18.                         **then** tc is included in the effective set of test cases;
19.                     **Endif**
20.                 **Endif**
21.             **Endfor**
22.         **Endfor**
23.     **Endfor**
24. **Endfor**

The resulting graph *TGout* = (*V tout*, *Edout* ) contains a vertex v = ($t_1$, ..., $tk$) for each transition sequence of length k. Two vertices v = ($t_1$, ..., $t_k$) and v′ = (t′$_1$, ..., t′$_k$) are connected by a directed edge if it holds that ($t_2$, ..., $t_k$) equals (t′$_1$, ..., t′$_{k-1}$) for each component. These two vertices thus represent the sequence ($t_1$, ..., $t_k$, t′$_k$).

# 6. WORKING OF OUR ALGORITHM

In this section, we explain the working of our algorithm using an automatic ticket machine example as described below.

The ATM(Automatic Ticket machine)issues tickets to passengers on receiving money from them. The state chart diagram of an automatic ticket machine, ATM object for various events of interest are shown in Fig 3. The object enters into *idle* state, when the power switch is on. Once the user selects an ticket type button in the menu, the object enters the *display destination* state, where the ticket fare to different destinations are displayed. The user can select the destination, ticket type and the number of persons(n) to travel. The condition n≤6 is inserted for the event *tickettypeselected*, as the ticket machine is not expected to issue a ticket for more than 6 persons in one transaction. Once the ticket type and number of persons required are selected, the object enters the *display money* state. In this state, the object displays the amount of money (totalfair) the user has to enter into the ticket machine. Note that totalfair = ticket fare × number of persons. As the user enters money (amt) into the machine, the machine object changes its state to busy. In the busy state, it calculates how much balance or change (chng) has to be returned to the user if any, where chng = (amt - totalfair). If the change balance is less than zero, the machine object changes its state from *busy* to *return money* as the money inserted is insufficient. If the change balance is more than or equal to zero, the machine object goes to *issueticket* state and issues the ticket for requested number of persons. If the balance is zero, then once the ticket is issued the machine object changes its state from *issueticket* to *idle*. If the balance (chng) is more than zero, the machine object enters into the *return money* state, where the change balance money is returned. Once the money is returned, the machine object again enters into *idle* state. Here the state chart diagram of the machine is given in the figure 3.

After constructing state chart diagram, we construct transition graph using algorithm-1. The transition graph is shown in figure 4. In next step, we extract all required information from the state transition graph as described below.

ST = {St1, St2, St3, St4, St5, St6, St7}, each $St_i$ is a stage or node

ID = {ID1,ID2,ID3,ID4,ID5,ID6,ID7,OD8,OD9}, each $ID_i$ is an input data

OD = {OD1, OD2, OD3, OD4, OD5, OD6, OD7, OD8, OD9}, each $OD_i$ is an output value

TR = {TR1,TR2,TR3,TR4,TR5,TR6,TR7}, each $TR_i$ is a transition between source and destination stage, where each $TR_i$ = {Stp, Stq}, Stp is a source stage and Stq is a destination stage. Hence, each transition can be extracted as follows:

TR1 = {St1, St2}

TR2 = {St2, St3}

TR3 = {St3, St4}

TR4 = {St4, St5}

TR5 = {St5, St6}

TR6 = {St5, St7}

TR7 = {St7, St6}

This step is to verify the completion of extracted information, derived from the diagram. Next, we derive and generate test cases. Hence, all tests can be generated as follows:

tc1 = {St1, St2, ID1, OD1,TR1}

tc2 = {St1, St2, St3, ID1, ID2, OD1, OD2, TR1, TR2}

tc3 = {St1, St2, St3, St4, ID1, ID2, ID3, OD1, OD2, OD3, TR1, TR2, TR3}

tc4 = {St1, St2, St3, St4, St5, ID1,..ID4, OD1,..OD4, TR1, ..TR4}

tc5 = {St1, St2, St3, St4, St5, St6, ID1,..ID5, OD1,..OD5, TR1, ..TR5}

tc6 = {St1, St2, St3, St4, St5, St7, ID1,..ID4, ID6, OD1,..OD4, OD6, TR1, ..TR4, TR6}

tc7 = {St1, St2, St3, St4, St5, St7, St6, ID1,..ID4, ID6, ID7, OD1,..OD4, OD6, OD7, TR1, TR6, TR7}

tc8 = {St2, St3, ID2, OD2,TR2}

tc9 = {St2, St3, St4, ID2, ID3, OD2, OD3, TR2, TR3}

tc10 = {St2, St3, St4, St5, ID2,..ID4, OD2,..OD4, TR2, ..TR4}

tc11 = {St2, St3, St4, St5, St6, ID2,..ID5, OD2,..OD5, TR2, ..TR5}

tc12 = {St2, St3, St4, St5, St7, ID2,..ID4, ID6, OD2,..OD4, OD6, TR2, ..TR4, TR6}

tc13 = {St2, St3, St4, St5, St7, St6, ID2,..ID4, ID6, ID7, OD2,..OD4, OD6, OD7, TR2, ..TR4, TR6, TR7}

tc14 = {St3, St4, ID3, OD3,TR3}

tc15 = {St3, St4, St5, ID3, ID4, OD3, OD4, TR3, TR4}

tc16 = {St3, St4, St5, St6, ID3,..ID5, OD3,..OD5, TR3, ..TR5}

tc17 = {St3, St4, St5, St7, ID3, ID4, ID6, OD3, OD4, OD6, TR3, TR4, TR6}

tc18 = {St3, St4, St5, St7, St6, ID3, ID4, ID6, ID7, OD3, OD4, OD6, OD7, TR3, TR4, TR6, TR7}

tc19 = {St4, St5, ID4, OD4,TR4}

tc20 = {St4, St5, St6, ID4, ID5, OD4, OD5, TR4, TR5}

tc21 = {St4, St5, St7, ID4, ID6, OD4, OD6, TR4, TR6}

tc22 = {St4, St5, St7, St6, ID4, ID6, ID7, OD4, OD6, OD7, TR4, TR6, TR7}

tc23 = {St5, St6, ID5, OD5,TR5}

tc24 = {St5, St7, ID6, OD6,TR6}

tc25 = {St5, St7, St6, ID6, ID7, OD6, OD7, TR6, TR7}

tc26 = {St7, St6, ID7, OD7,TR7}

The last step is to minimize the set of test cases by calculating node coverage for each test case and determine which test cases are covered by other test cases.

NC(tc1) = {tc2, tc3, tc4, tc5, tc6, tc7}

NC(tc2) = {tc3, tc4, tc5, tc6, tc7}

NC(tc3) = {tc4, tc5, tc6, tc7}

NC(tc4) = {tc5, tc6, tc7}

NC(tc5) = { }

NC(tc6) = {tc7}

NC(tc7) = { }

NC(tc8) = {tc9, tc10, tc11, tc12, tc13}

NC(tc9) = {tc10, tc11, tc12, tc13}

NC(tc10) = {tc11, tc12, tc13}

NC(tc11) = {tc13}

NC(tc12) = {tc13}

NC(tc13) = { }

NC(tc14) = {tc15, tc16, tc17, tc18}

NC(tc15) = {tc16, tc17, tc18}

NC(tc16) = {tc18}

NC(tc17) = {tc18}

NC(tc18) = { }

NC(tc19) = {tc20, tc21, tc22}

NC(tc20) = {tc22}

NC(tc21) = {tc22}

NC(tc22) = { }

NC(tc23) = {tc25}

NC(tc24) = {tc25}

NC(tc25) = { }

NC(tc26) = {tc25, tc22, tc18, tc13, tc7}

Therefore, the following test cases like tc1, tc2, tc3, tc4, tc6, tc8, tc9, tc10, tc11, tc12, tc14, tc15, tc16, tc17, tc19, tc20, tc21, tc23, tc24, tc26 should be ignored. Hence, the remaining effective set of test cases is TS = {tc5, tc7, tc13, tc18, tc22, tc25}.

**6.1. Reduction**
In practical situations, there is often insufficient time for thorough testing activities within industrial projects. Hence, it is reasonable to try to reduce the size of generated test suites. However, the effect of the reduction on the fault-detection ability of the test suites should be small. The techniques proposed in this paper can be used to apply reduction during test case generation. A single test case may cover more than the coverage item it has been generated for. When using a probe based technique as described in this paper it is easy to identify all items covered by a particular test case.

# 7. RELATED WORK

A lot of studies have investigated the effect of test-set reduction on the size and fault finding capability of a test set.

In an early study, Wong et al. [27 ]address the question of the effect on fault detection of reducing the size of a test set while holding coverage constant [27], [28]. They randomly generated a large collection of test sets that achieved block and all-uses data flow coverage for each subject program. For each test set they created a minimal subset that preserved the coverage of the original set. They then compared the fault finding capability of the reduced test-set to that of the original set. Their data shows that test minimization keeping coverage constant results in little or no reduction in its fault detection effectiveness. This observation leads to the conclusion that test cases that do not contribute to additional coverage are likely to be ineffective in detecting additional faults.

To confirm or refute the results in the Wong study, Rothermel et al. [24] performed a similar experiment using seven sets of C programs with manually seeded faults [24]. For their experiment they used edge-coverage [7] adequate test suites containing redundant tests and compared the fault finding of the reduced sets to the full test sets. In this experiment, they found that [2] the fault-finding capability was significantly compromised when the test-sets were reduced and [25] there was little correlation between test-set size and fault finding capability. The results of the Rothermel [24] study were also observed by Jones and Harrold in a similar experiment [10].

Offutt and Abdurazik [19], [20] developed a technique for generating test cases from UML state diagrams. They have highlighted several useful test coverage criteria for UML state charts such as: (1) full predicate coverage, (2) transition coverage etc.

Kansomkeat and Rivepiboon [11] have introduced a method for generating test sequences using UML state chart diagrams. They transformed the state chart diagram into a flattened structure of states called testing flow graph (TFG). From the TFG, they listed possible event sequences which they considered as test sequences. The testing criterion they used to guide the generation of test sequences is the coverage of the states and transitions of TFG.

Kim et al. [12] proposed a method for generating test cases for class testing using UML state chart diagrams. They transformed state charts to extended FSMs (EFSMs) to derive test cases. In the resulting EFSMs, the hierarchical and concurrent structure of states are flattened and broadcast communications are eliminated. Then data flow is identified by transforming the EFSMs into flow graphs, to which conventional data flow analysis techniques are applied. These different results are difficult to reconcile and the relationship between coverage criteria, test-suite size, and fault finding capability clearly needs more study. In the experiment discussed in this paper we attempt to highlight some additional issue. Our work is different in some respects. First, we are not studying testing of traditional programs, we are interested in test-case generation, test case minimization by calculating node coverage for each test case and testing of specifications. A single test-case in most cases may satisfy more than one test obligation. For instance, a test case used to cover a certain state of interest may also cover other states during its execution. It provides a way to reduce the size of the final test-suite by choosing a subset of test-cases that preserves the coverage obtained by the full test-suite.

## 8. CONCLUSSION AND FUTURE WORK

In this paper, first, we build a state chart model of our system under test. Next, we derived state transition graph from state chart diagram. Then, all required information are extracted from the graph. Then, the test cases are generated by applying Wang's algorithm. Lastly, a set of test cases are minimized by calculating node coverage for each test case and it is determined that which test case are covered by other test cases. In this way, this paper introduces efficient test generation technique to optimize test coverage by minimizing time and cost.

This paper presents a coverage and specification-oriented test approach based on state chart diagram. In our opinion, there is an unacceptable loss in terms of test-suite quality. Thus, we advocate research into test-case prioritization techniques and experimental studies to determine if such techniques can more reliably lessen the burden of the testing effort by running a subset of an ordered test suite as opposed to a reduced test suite, without loss in fault finding capability. Further research is planned to extend the model for considering time constraints to handle more intricate applications.

## REFERENCES

[1]   OMG. Unified Modeling Language specification, version 2.0, August 2005, object management group, www.omg.org.

[2]   M. Archer, C. Heitmeyer, and S. Sims, 1998. *TAME: A PVS interface to simplify proofs for automata models, In User Interfaces for Theorem Provers*.

[3]   Fevzi Belli and Axel Hollmann, March 2008. Test generation and minimization with basic statecharts. *ACM, SAC-08*, pages 718 – 723.

[4]   Michel R. Blaha and James R. Rumbaugh. *Object-Oriented Modeling and Design with UML*. Pearson, second edition.

[5]   M. Broy, June 2005. Model-based testing of reactive systems. Advanced Lectures , Springer..

[6]   T. S. Chow, 1978. Testing software design modeled by finite-state machines. *IEEE TSE*, 4(3): , pages 178 – 187.

[7]   P. Frankl and S. N. Weiss, 1991. An experimental comparison of the effectiveness of the all-uses and all-edges adequacy criteria. In *Proceedings of the symposium on Testing, analysis, and verification*.

[8]     D. Harel, 1987.. Statecharts: A visual formulation for complex systems. *Sci. Comp. Prog.*, 8: , pages 231 – 274

[9]     Mats P. E. Heimdahl and Devaraj George. Test-suite reduction for model based tests: Effects on test quality and implications for testing.

[10]    J. A. Jones and M. J. Harrold, March 2003. Test-suite reduction and prioritization for modified condition/decision coverage. *IEEE Transactions on Software Emgineering*, 29(3): , pages 195 – 209.

[11]    S. Kansomkeat and W. Rivepiboon, 2003. Automated-generating test case using UML statechart diagrams. In *Proc. SAICSIT 2003, ACM*, pages 296 – 300.

[12]    Y. G. Kim, H. S. Hong, D. H. Bae, and S. D. Cha et a, 1999l. *Test cases generation from UML state diagram, Software Testing Verification and Reliability,* , pages 187 – 192.

[13]    Nicha Kosindrecha and Jirapun Daengdej, 2005 – 2010. A test generation method based on state diagram. *journal of Theoritical and Applied Information Technology*, pages 28 – 44.

[14]    R. Lai, 2002. A survey of communication protocol testing. *Journal of Systems and Software*, 62(1): , pages 21 – 46.

[15]    J. T. Lallchandani and R. Mall, 2010. Integrated state-based dynamic slicing technique for UML models. In *IET Software, Vol. 4, Iss. 1*, pages 55 –78.

[16]    H. Li and L. C. Peng, January 2005. Software test data generation using ant colony optimization. In *Proceedings of World Academy of Science, Engineeing and Technology*.

[17]    Wang Linzhang, Yuan Jiesong, Yu Xiaofeng, Hu Ju, Li Xuandong, and Zheng Guoliang, 2004. Generating test cases from UML activity diagram based on gray-box method. *Proceedings of the 11th Asia-Pacific Software Engineering Conference (APSEC04)*.

[18]    R. Mall, 2009. *Fundamentals of Software Engineering*. Prentice Hall, 3$^{rd}$ edition.

[19]    J. Offutt and A. Abdurazik, 1999. Generating tests from uml specifications. In *Proceedings of 2nd International Conference. UML, Lecture Notes in Computer Science*, pages 416 – 429.

[20]    J. Offutt, S. Liu, A. Abdurazik, and P. Ammann et al. , 2003 Generating test data from state-based specifications. *Software Testing Verification Reliability.*, 13: , pages 25 – 53.

[21]    D. Pilone and N. Pitman, 2005. *UML 2.0 in a Nutshell*. NY. O'Reilly, USA.

[22]    M. Priestley. *Practical Object-Oriented Design with UML*. Tata McGraw-Hill, second edition.

[23]    G. Reinelt, 1994. In the traveling salesman: Computational solutions for tsp applications. *Springer Berlin / Heidelberg*, 840.

[24]    G. Rothermel, M. Harrold, J. Ostrin, and C. Hong, November 1998. An empirical study of the effects of minimization on the fault detection capabilities of test suites. In *Proceedings of the International Conference on Software Maintenance*, pages 34 – 43.

[25]    C. Parent-Vigouroux S. Bensalem, P. Caspi and C. Dumas, 1999. A methodology for proving control systems with lustre and pvs. In *In Proceedings of the Seventh Working Conference on Dependable Computing for Critical Applications (DCCA 7)*.

[26]    M. Sharma and R. Mall, 2009. Automatic generation of test specifications for coverage of system state transitions. *Information and Software Technology*, (51): , pages 418 – 432.

[27]    W. Wong, J. Horgan, S. London, and A. Mathur, April 1998. Effect of test set minimization on fault detection effectiveness. *Software Practice and Experience*, 28(4): , pages 347 – 369.

[28]    W. Wong, J. Horgan, A. Mathur, and A. Pasquini, August 1997. Test set size minimization and fault detection effectiveness: A case study in a space application. In *Proceedings of the 21st Annual International Computer Software and Applications Conference*, pages 522 – 528.

**Authors**

**Ranjita Ku. Swain** completed her MCA from College of Engg.And Technology, OUAT, Bhubaneswar, India. She is pursuing her Ph.D degree from Utkal University, Vani vihar, Bhubaneswar, India. She is currently working as Senior Lecturer in Computer Science Dept., Rourkela Institute of Management Studies, Rourkela, India. She has 11years of teaching experience and her fields of interest are Software Engg., Discrete, Mathematical Structure and Numerical Methods.

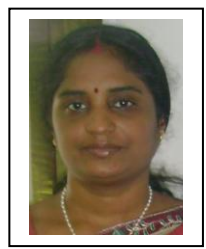

**Prafulla Ku. Behera** has received his Ph.D degree from Utkal University, Vani vihar, Bhubaneswar, India. He is currently working as a reader at Dept. of Computer Science & Application, in Utkal University, Vani vihar, Bhubaneswar, India. His special fields of interest include Mobile Computing, Software Engineering, He is a member of CSI.

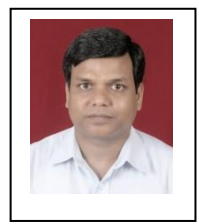

**Durga Prasad Mohapatra** received his Ph. D. from Indian Institute of Technology Kharagpur and M. E. from Regional Engineering College (now NIT), Rourkela. He joined the faculty of the Department of Computer Science and Engineering at the National Institute of Technology, Rourkela in 1996, where he is now Associate Professor. His research interests include software engineering, real-time systems, discrete mathematics and distributed computing and published more than forty papers in these fields. He has received many awards including Young Scientist Award for the year 2006 by Orissa Bigyan Academy, Prof. K. Arumugam award for innovative research for the year 2009 and Maharasthra State National Award for outstanding research for the year 2010 by ISTE, NewDelhi. He has also received three research projects from DST and UGC. Currently, he is a member of IEEE. Dr. Mohapatra has co-authored the book *Elements of Discrete Mathematics: A computer Oriented Approach* published by Tata Mc-Graw Hill.

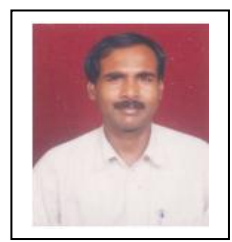